# Electronic Medication Prescribing Support System for Diagnosing Tropical Diseases

[1]Omotosho, A, [2]Olaniyi, O.M, [3]Emuoyibofarhe, O.J, [4]Osobu, O.B.
[1,4]Department of Computer Science and Technology
[1,4]Bells University of Technology, Ota , Ogun-state, Nigeria.
[2]Department of Computer Engineering
[2]Federal University of Technology, Minna ,Niger-state, Nigeria
[3]Department of Computer Science and Engineering
[3]Ladoke Akintola University of Technology, Ogbomoso , Oyo State, Nigeria
E-mail [1]bayosite2000@yahoo.com [2]engrolaniyi09@yahoo.com

**ABSTRACT**

Tropical diseases are diseases that are unique to tropical and subtropical regions affecting nearly everyone in the "bottom billion" of the world's poorest people. They have a terrible impact on health involving child growth and development, harm pregnant women, cause disability and disfigurement and often cause long-term debilitating illnesses. The conventional mode of medical prescription by doctors is usually handwritten and hand-delivered by patient to the dispensing pharmacist. This method has been found to be prone to prescription errors and it is an inefficient means to store and access a patients' prescription record for future decisions on patient health. Majority of the flaws inherent in the conventional approach arises from fairly illegible handwriting of the physician, misinterpretation of a prescription by the pharmacist, lack of knowledge of a patient's health record, and also the queue experienced by the patient when it comes to collecting drugs. This paper presents the development of an e-prescription system for diagnosing tropical diseases.Results after testing the developed system by medical experts indicated that the e-prescription systems is more efficient and less susceptible to common errors associated with the conventional handwritten medical prescription and can also go a long way to help to improve patients health outcome in the health industry especially in the tropics.

## 1	INTRODUCTION

Tropical diseases are associated with a high level of mortality rate and are also very common in developing countries. The tropical diseases under consideration in this study are malaria, typhoid and tuberculosis. Malaria is an important cause of death and illness in children and adults, especially in tropical countries (WHO, 2010). It thrives in the tropical areas of Asia, Africa, Central and South America, where it strikes millions of people. Each year 350 to 500 million cases of malaria occur worldwide. Unfortunately, more than 1 million of its victims, mostly young children, die yearly (NIAID, 2007). In endemic areas and in large outbreaks, most cases of Typhoid Fever occur in persons aged between 3 and 19 years. In 1997, for example, this age range was reported during an epidemic of the disease in Tajikistan. Nevertheless, clinically apparent infection in children aged less than three years has been described in Bangladesh, India, Jordan, Nigeria, and elsewhere (WHO, 2003). In Indonesia, people aged 3-19 years accounted for 91% of cases of Typhoid Fever and the attack rate of blood-culture-positive Typhoid Fever was 1026 per 100 000 per year. (WHO, 2003). Between 1% and 5% of patients with acute Typhoid infection have been reported to become chronic carriers of the infection in the gall bladder, depending on age, sex and treatment regimen (WHO, 2003). Tuberculosis (TB) is the second-largest cause of death from an infectious agent worldwide—killing approximately 1.7 million people in 2003. Despite steady drops in the number of cases in some parts of the world, the number of new cases appears to be growing, with an estimated 8.8 million new cases in 2003 (DCPP, 2006). Sub-Saharan Africa has the highest incidence rate at 345 new cases per 100,000 inhabitants per year, while one-half of the world's cases are in Bangladesh, China, India, Indonesia, and Pakistan. Due to suppressed immune systems, people living with HIV/AIDS are more likely to become infected with TB. TB is a major cause of death among those with HIV/AIDS (Varaine, 2010).

Over the last decade, the need to develop and organize new ways of providing efficient health-care services has been accompanied by major advances in information and communications technology (ICT). This has resulted in a dramatic increase in the use of ICT applications in health care, collectively known as e-health. E-Health is the use, in the health sector, of digital data -transmitted, stored and retrieved electronically- in support of health care, both at the local site and at a distance (WHO, 2007). The orthodox approach to medication prescription is the paper method which has been found to be prone to many errors which may aggravate patients' health. Such errors include the not so legible handwritten prescription which may cause delay in the dispensing process, misinterpretation of medication prescription, drug-to-drug reaction, and drug to allergy interaction (Omotosho and Emuoyibofarhe, 2012). The paper method of prescription in addition, causes stress to the patient as they would have to hand-deliver the paper prescription to the pharmacist even in the ill state of the patient and worse would have to wait on queue. But, with the use of the e-prescription system – medical drug prescription would be composed on a system by the prescribing physician by selecting the drug name, accessing the patient record and then sending electronically to the pharmacy. In e-prescribing system, the pharmacist could prepare the medication which would be ready for receipt before the patient arrives. Failures in communication between the pharmacist and the prescriber can also result in error in far too many cases besides the handwriting of the physician (Miller et al, 2005). This process of the physician sending his prescription electronically also reduces the rate of call-backs to and from the pharmacy for clarity of prescription.

Research and evidences have shown that medication errors are a well-known problem in hospitals and that medication errors and adverse drug reactions (ADRs) are one of the main causes of hospitalization leading to disability and even death in up to 6.5% of hospital admissions (Emily et al, 2005). The errors caused by the inherent problems of the manual method are potentially correctible and reduced at the point of the prescriber (Physician) with the advent of technologies such as e-prescription which automates the process of prescription and with the application of a decision support system associated with drug knowledge and dissemination.

This paper presents the development of an electronic prescription system for selected tropical diseases and it is expected to improve healthcare delivery when providing medication to tropical diseases in the tropics.

## 2      RELATED WORK

Electronic prescription is the use of information and communication technologies (ICT) and tools to acquire, examine, re-examine, modify, and electronically transmit prescription information about pharmaceutical products by legally and professionally qualified and registered healthcare practitioners to registered pharmacies (or dispensing systems). (Omotosho and Emuyibofarhe,2011). The main prescribing practice in majority of the third world countries- the paper method, offers weakness in the delivery of high quality medical care. In general, medical prescription errors are caused primarily by communication of prescriptions, illegible handwriting, unclear abbreviations, dose errors, unclear oral orders, ambiguous orders and fax clarity. An Electronic prescribing (e-prescribing) would require typed input or a selection from a drop-down menu, preferably from the physician, which would avoid most of these errors or allow them to be rectified online by the physician at the time of prescription writing. (Lyman, 2007). In third world countries like Peru,  Egypt  and  Uganda, effective  use  of  Information Communication Technologies  has  prevented  avoidable  maternal  deaths  and therefore the prospect of ICT in health cannot be overestimated (Omotosho et al., 2011) .

Although the concept of computerization of medical prescription process is still developing however there are few functional and existing system already. Electronic prescription is currently widely in use especially in the some countries, like Sweden, Finland, Denmark and also in the Netherlands. Emily, (2005), discussed the development and implementation of a Computerized physician order entry (CPOE) system focusing specifically on a home-grown e-prescribing system, in a community-based, integrated health care system. This was achieved by ensuring the buy-in of the head of the organisation, ensuring a two way communication, iterative implementation, ongoing and readily accessible training to be available, the involvements of clinicians in all facets of development that contributes to improvements, and also the workflow be redesigned. From the article, if their theories are implemented practically as specified then it would be an almost perfect implementation. Rahimi at al, (2009), analysed the integrated electronic prescribing systems (IEPSs) used in Sweden. In this system, every e-prescribing hub and pharmacy are interconnected and can send and receive prescriptions electronically. In Sweden, every newly born child is given an identifier number in which is used as an identifier for that citizen whenever his patient record needs to be accessed or identified. They improved efficiency and safety in the management of pharmaceuticals throughout the healthcare sector, smoother prescribing, better service to the patients and time saving for all parties. ICT tools promises to improve the processes in the health industry as a whole, Olabiyisi et al, (2011), used fuzzy set theory and fuzzy logic to develop a knowledge-based system in medicine for diagnostic task. Their design was basically to demonstrate how information technology and medicine can successfully operate together using differential diagnosis by applying fuzzy logic to medical informatics. they verified their approach with trials on systems such as case based reasoning for medical diagnosis using fuzzy set theory and have shown that fuzzy logic can resolve the conflict arising from ambiguity, uncertainty and imprecision of information. Omotosho and Emuoyibofarhe, (2012), proposed and developed a model for secure intelligent e-prescription system using fingerprint based biometric for authentication and also with an intelligent database that automatically validate prescription. Therefore, reduces the incidences of prescription error. In this model, the prescriber can access a patient record at an external source including the national database. However, the system was not designed to handle specific diseases.

This work presents an electronic prescription system designed to handle specific tropical diseases.

## 3  MATERIALS AND METHODOLOGY

The following scientific approaches were used to achieve the central idea of this work. They are: Requirement definition and infrastructural modeling.

### 3.1  Requirement Definition of the Proposed Service Infrastructure

I. The e-prescription system for diagnose tropical diseases System Requirement

This requirement follows from the assumption that in order to automate the medical prescription process for tropical diseases, the system should provide:  a) **Eligibility and Authentication:** The system should be designed in a way that only allows access to authorized personnel. b) **Uniqueness:** A prescriber has only one login credential and it can not be used by another person. c) **Accuracy:** The administrator and prescribing physician should be able to compute records and generate patient prescription reports with lesser errors. d) **Integrity:**  patients' prescription records can only be modified, updated or deleted by the assigned administrative physician. e) **Reliability:** The system should work robustly without any loss of records due to good and reliable database and also should be able to deliver prescription to pharmacist in lesser time. f) **Flexibility:** More modules expected of medical prescription operations can be integrated into the system to increase functionality. g) **Convenience**: prescribing physician should be able to complete/cancel prescription process in minimal time. Pharmacists have no need for call-backs to physicians; Physicians have easy and direct access to a patient record and also the patient gets drugs from pharmacists without having to queue for them.

### 3.2  Infrastructural Model and Architect

i. **Overall System Architecture**

The system architecture of the designed system is shown in Figure 1; it illustrates how the system operates and functions. It consists of an application server that authenticates and verifies the logon on credentials by both the physician and the administrator.  An authentication function is built into the system via a username, in this case the practicing physician's license number and a password that verifies the users who log onto the system, a hospital database server where patient data, drug data are stored, the prescription database where the prescribed drug is sent to and picked up by the pharmacist to be dispensed to the patient.

The entire developed system operates in such a way that when a registered patient visits the hospital, after consultation with the physician and the patient has been diagnosed for a particular disease, the physician then logs onto the e-prescription system. The doctors' credential is verified by the application server that runs as an application to the computer. Prior to making prescription by the doctor, the prescribing physician can checks through the hospital Electronic Health Record for the patients' record for the most recent medications prescribed to the patient, his allergies and drug to drug interactions if any.

After that has been confirmed, the doctor then may proceed to composes his prescription and sends to the pharmacy. At then pharmacist end, a logon is made to a similar electronic pharmacy application; the pharmacist can then access the sent prescription residing in the prescription database before or on the patient arrival. Then e-prescription on receipt by the pharmacist is read and electronically acknowledged. He selects the drug to be dispensed based on the prescription made by the prescriber, once the drugs are gathered together in a single pack and ready for dispensation. The pharmacist guides the patient on how to use the drug and then dispenses it alternatively, the pharmacist may give a printed copy of the prescription to the patient should he decides to collect it from another pharmacy.

The Application software allows the users of the system, i.e. the administrator, the physician and the pharmacists to perform the basic circulation operations, i.e. add patients, prescribe to patients, looking up patient medical history, checking of prescription by pharmacist to be dispensed to the patient and making of drug formulary by the physicians'. Also is the ability to more diseases and to add more users (Pharmacists and Physician) to access the system (only the Administrator can add more users of the system).

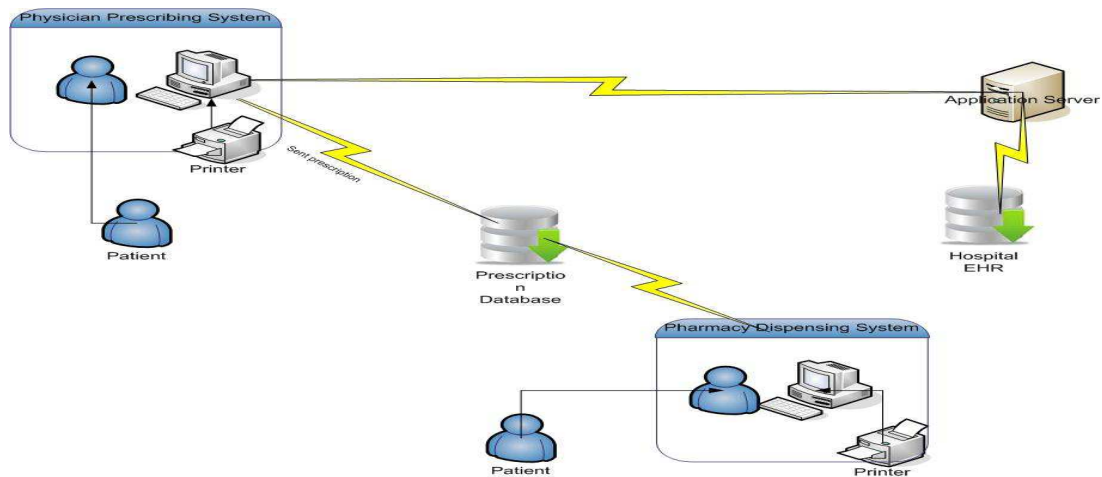

Figure 1: The System Architecture of the Designed E-prescription System

In designing the system the development tools employed includes: PHP (Hypertext Pre-Processor) which is a general-purpose server-side scripting language designed for web development to produce dynamic web pages, MySQL which is an open source database which primarily is a relational database management system (RDBMS) used commonly for web applications.

**ii System Modeling and Model Representation**

The software phase is divided into 3 sub-phases: Presentation/Front End (the application which interfaces the users would interact with), Logic/Middle End (dynamic content processing and generation level application server), Data/Back End (the database where the information is stored). Thus, the system architecture is a three-3 tier architecture which is an architectural deployment style that describe the separation of functionality into layers with each segment being a tier that can be located on a physically separate computer.

Using the standard Unified Modeling Language (UML), the designed Electronic Prescribing Support System for Diagnosing Tropical Diseases system was visualized along the following use case, class diagram, sequence, and activity diagrams. Figure 2 shows the use case diagram of the designed system. The functionality of Electronic Prescribing Support System for Diagnosing Tropical Diseases system is represented in terms of actors, their goals represented as use cases and available dependencies. Figure 3 shows the class diagram of the electronic prescribing system, Figure 4 shows the sequence diagram of the electronic prescribing system, Figure 5 shows the activity diagram for pharmacist and doctor and figure 6 shows the activity diagram of the electronic prescribing system administrator.

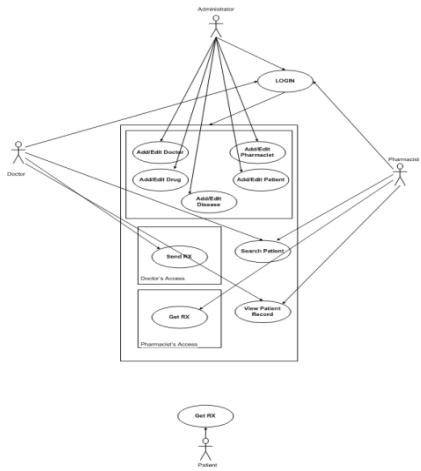

Figure 2: Use-Case Diagram of the Electronic Prescribing System

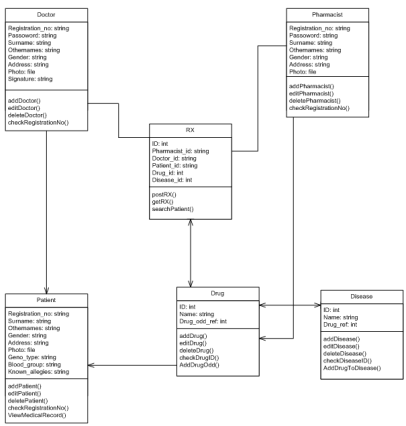

Figure 3: Class Diagram of the Electronic Prescribing System

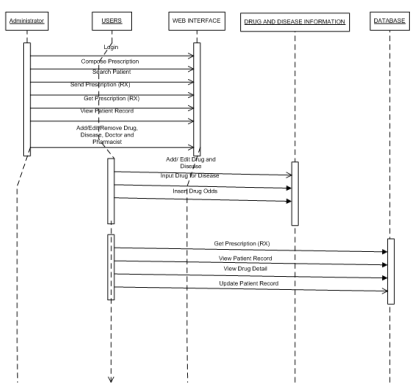

Figure 4: Sequence Diagram of the Electronic Prescribing System

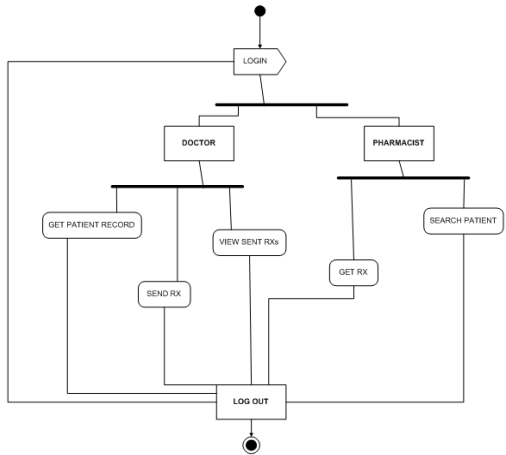

Figure 5: Activity Diagram for Pharmacist and Doctor

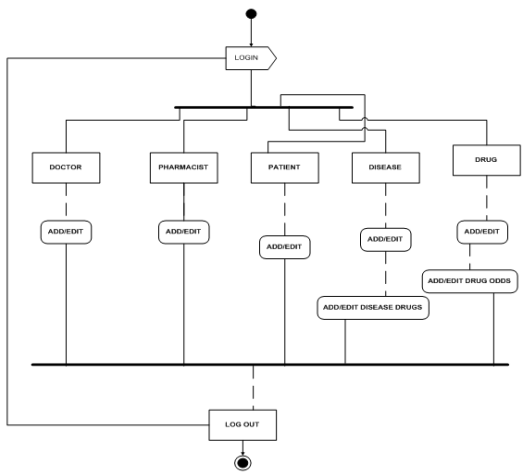

Figure 6: Activity Diagram of the Electronic Prescribing System Administrator

## 4      SYSTEM IMPLEMENTATION

The system implementation covers a broad spectrum of activities occurred within the system and they are depicted using snapshot of functional key areas of the system including;

**Login Page**

The system only allows authorized users including the administrator, doctor, and pharmacist to access it in Figure 7. The administrator who is the user with higher privilege to add, remove and edit features such as drugs, diseases to the system including the creation and addition of users. He is the user responsible for maintaining the entire system – ensure it is up and running at all times. Also are the doctor and pharmacist, who make and dispense prescription respectively.

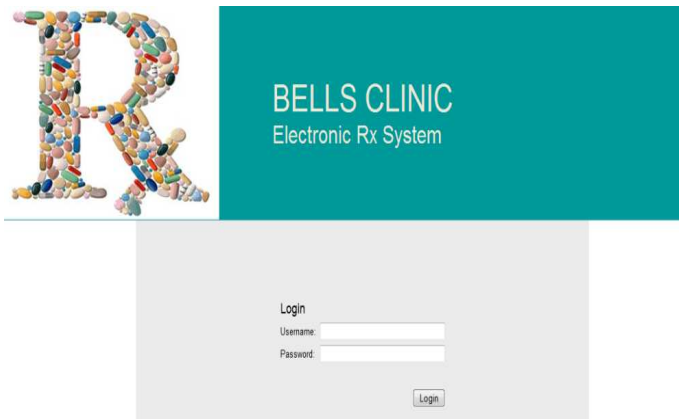

Figure 7: Login Page

**Administrators' Page**

Through the interface in figure 8, functions to add, remove and edit the doctor, pharmacist, patient, drug and disease are available. The interface is only accessible to the administrator after he must have logged in with his original and registered credentials. After the login credentials of the administrator have been authenticated, the system opens up with several tabs - these tabs allow the administrator to make changes in the database relating to the doctors, pharmacists, disease and drug. This tab is inaccessible to the doctors and pharmacist.

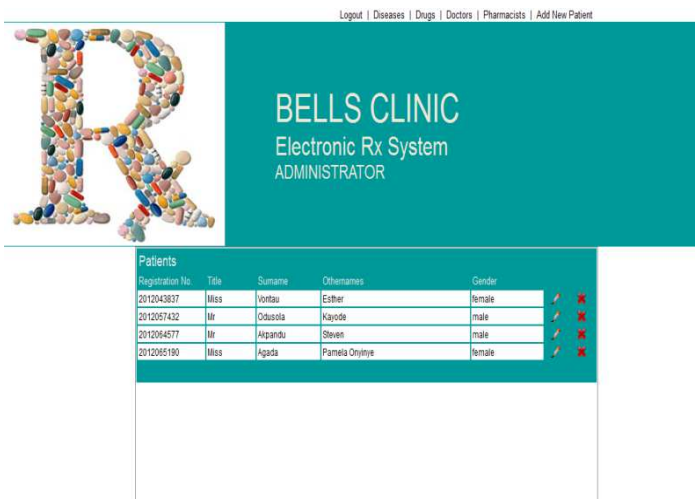

Figure 8: Administrator Page

**Doctors' Page**

In figure 9, the doctor who is logged onto the system has the ability to make prescription by clicking the "Compose RX" tab. He can also search for patients, view a patients' record, the drug detail and can send the prescription to the pharmacy.

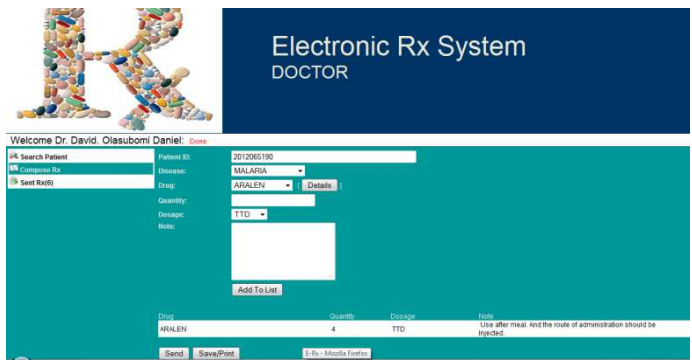

Figure 9: Prescription page of doctor

**Pharmacist Page**

In figure 10, the pharmacist is logged onto the system. He can view pending prescriptions, search for a patient, and also view the drug detail. Then afterwards would deal the prescription by clicking process.

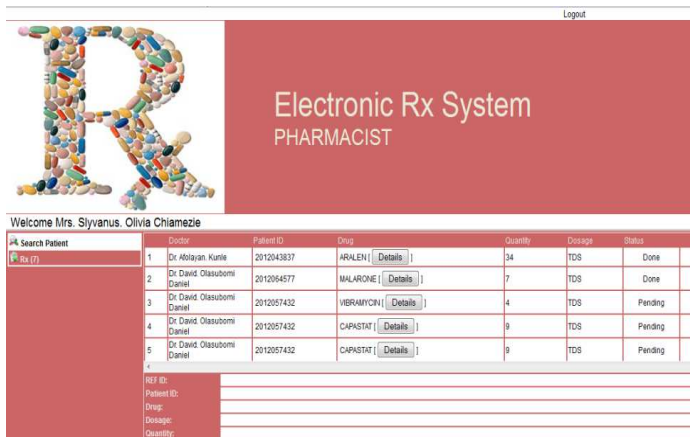

Figure 10: Pharmacist page

**Drug Details**

In figure 11, both the doctor and pharmacist when logged onto the system can view the drug detail before making a prescription by the former and before dispensing also by the latter. This detail contains the pharmaceutical class of the drug, generic description, indications, and adverse reactions.

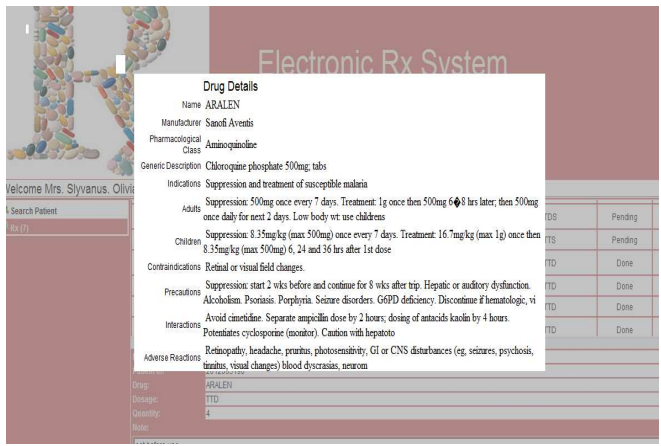

Figure 11: The detail of drug

## 5  CONCLUSION

This paper has successfully presented the development of electronic prescribing support system for diagnosing tropical diseases due to the inherent problems in the conventional method of medical prescription in the tropics. These problems range from ambiguous writing style by doctors, difficulty in accessing patient record, the unreliable method in keeping medical record and also the queue posed on patients when coming to get the prescribed drug. We therefore recommended the deployment of the system at the national level where patient record can be centrally stored and accessed. Also, the system can be made interoperable with mobile devices. Medical practitioners should be trained on the use of the system and encouraged of the advantages posed when using the system. In future the e-prescription system shall be evaluated by medical practitioners to establish the level of satisfactory on health outcomes.